\documentstyle[prl,aps,preprint,tighten,floats,psfig]{revtex}

\begin{document}

\draft

\preprint{\vbox{\hfill UMN-D-02-1  \\
          \vbox{\hfill hep-th/0203162}
          \vbox{\vskip0.3in}
          }}

\title{Approximate BPS States}

\author{J.R. Hiller}
\address{Department of Physics, 
University of Minnesota Duluth, 
Duluth, MN 55812}

\author{S.S. Pinsky and U. Trittmann}
\address{Department of Physics, 
The Ohio State University, 
Columbus, OH 43210}

\date{\today}

\maketitle

\begin{abstract}
We consider dimensionally reduced three-dimensional supersymmetric
Yang--Mills-Chern--Simons theory. 
Although the ${\cal N}=1$ supersymmetry of this theory does not allow 
true massive Bogomol'nyi--Prasad--Sommerfield (BPS) states, 
we find approximate BPS states which have non-zero masses that are 
almost independent of the Yang--Mills coupling constant and which 
are a reflection of the massless BPS states of the underlying 
${\cal N}=1$ super Yang--Mills theory.  
The masses of these states at large Yang--Mills 
coupling are exactly at the $n$-particle continuum thresholds. 
This leads to a relation between their masses at zero and large 
Yang--Mills coupling.    
\end{abstract}

\narrowtext


Bogomol'nyi--Prasad--Sommerfield (BPS) states play an important role 
in modern quantum field theory for a variety of reasons. In particular,
their property of having masses which are independent of the coupling
is a very useful tool, because it allows the evaluation of the spectrum 
in the nonperturbative region where it is otherwise difficult if not 
impossible to solve the theory.  BPS saturated states arise because 
the supersymmetry protects the masses of states which are destroyed by 
a linear combination of the supercharges while forcing these masses to 
be equal to the central charge in appropriate units.  
It is well known that one must have at least
${\cal N}=2$ supersymmetry to have a non-zero central charge and therefore
massive BPS saturated states. Many of the more interesting field theories,
at least from the phenomenological point of view, only have ${\cal N}=1$
supersymmetry, and this magical property of BPS states is of no help in
calculating the massive spectrum in the nonperturbative regime.

We will show by explicit
demonstration that it is possible to find {\em approximate} BPS states in
theories with ${\cal N}=1$ supersymmetry. These are states with
masses that are to a very good approximation independent 
of the coupling but are not true BPS
states. We will show that the origin of these states is the presence of
massless BPS states in the closely related ${\cal N}=1$ supersymmetric
Yang--Mills (SYM) theories. These theories have BPS saturated states that are
massless and are destroyed by one of the supercharges.  

For this demonstration we consider dimensionally reduced SYM-Chern--Simons
theory.  This theory has the advantage that the partons are given a bare mass
without breaking the supersymmetry, which makes the theory particularly suitable 
for numerical studies.  However, such theories are also interesting in their own
right.  A Chern--Simons (CS) theory can be used to study many interesting
phenomena, such as~\cite{dunne} the quantum Hall effect, Landau levels,
non-trivial topological structures, vortices, and anyons. 
According to Witten~\cite{witten}, it is possible that string
field theory is essentially a non-commutative CS theory.  This idea
led to a conjecture by Susskind~\cite{susskind} that relates string theory
to the fractional quantum Hall effect.
The SYM-CS theories are particularly remarkable.  As is well known,
there is a finite anomaly that shifts the CS coupling~\cite{lee}.
Moreover, Witten~\cite{witten2} has conjectured that this theory spontaneously
breaks supersymmetry for some values of the CS coupling.  

In this Letter we briefly discuss SYM-CS theory in 2+1 dimensions but then
dimensionally reduce it to two dimensions by requiring all of the fields
to be independent of the transverse coordinate.  This reduction eliminates
many of the most interesting aspects of CS theory, including the quantization 
of the CS coupling, but does preserve the fact that the CS term
simulates a mass for the theory. The effective mass leads to QCD-like 
properties for the theory, which are discussed in~\cite{hpt01}.

The numerical method we use to solve the SYM-CS theory
is supersymmetric discrete light-cone quantization (SDLCQ).  This
method can be used to solve any theory with enough
supersymmetry to be finite.  
By use of ordinary discrete light-cone quantization (DLCQ)~\cite{pb85,bpp98}
we can construct a finite dimensional representation of the
superalgebra~\cite{sakai95}.  From this representation of the superalgebra,
we construct a finite-dimensional Hamiltonian which we diagonalize
numerically.  Unlike direct discretization of the Hamiltonian, this 
construction automatically preserves supersymmetry exactly.
We repeat the construction for larger and larger representations
and extrapolate the solution to the continuum.  We have already used
this method to solve (1+1) and (2+1)-dimensional SYM 
theories~\cite{alpt98,Haney:2000tk,Antonuccio:1999zu,Antonuccio:1998jg,%
Antonuccio:1998kz}~\footnote{For additional references to this work,
see~\cite{hpt01}.} as well as the dimensionally reduced SYM-CS theory~\cite{hpt01}.

To understand the properties of dimensionally reduced SYM-CS theory we
will need to review earlier results of similarly reduced SYM theory. 
This ${\cal N}=1$ SYM theory in 1+1 dimensions is a stringy theory, in the sense
that the low-mass states are dominated by Fock states with many constituents. 
As the size of the discrete superalgebra representation is increased, 
states with lower masses and more constituents
appear~\cite{Antonuccio:1998jg,Antonuccio:1998kz}.
In addition, this theory has a well-defined number of massless BPS states.



We begin by considering  ${\cal N}=1$ supersymmetric CS theory in 2+1
dimensions. The Lagrangian  of this theory is 
\begin{eqnarray} \label{Lagrangian}
{\cal L}&=&{\rm Tr}\left(-\frac{1}{4}F_{\mu\nu}F^{\mu\nu}
   +i\bar{\Psi}\gamma_{\mu}D^{\mu}\Psi+2\bar{\Psi}\Psi \right. \\
    &&\left.+\frac{\kappa}{2}\epsilon^{\mu\nu\lambda}\left[A_{\mu}
    \partial_{\nu}A_{\lambda}+\frac{2i}{3}gA_\mu A_\nu A_\lambda \right]
      \right)\,, \nonumber
\end{eqnarray} 
where $\kappa$ is the CS coupling.  The two components of the spinor
$\Psi=2^{-1/4}({\psi \atop \chi})$ are in the adjoint representation, and
we will work in the large-$N_c$ limit.  The field strength and the
covariant derivative are
$F_{\mu\nu}=\partial_{\mu}A_{\nu}-\partial_{\nu}A_{\mu}+ig[A_{\mu},A_{\nu}]$ 
and $D_{\mu}=\partial_{\mu}+ig[A_{\mu},\quad]$, respectively.
Supersymmetric variations of the fields lead to the supercharge components 
\begin{eqnarray} \label{supercharges} 
Q^-&=&-i 2^{3/4}\int d^2x\,
   \psi\left(\partial^+ A^- -\partial^-A^+ + ig[A^+,A^-]\right)\,, 
\nonumber \\ 
Q^+&=&-i 2^{5/4}\int d^2x\, 
  \psi \left(\partial^+ A^2 -\partial^2 A^+ +ig[A^+,A^2]\right)\,. 
\end{eqnarray} 
The supercharge fulfills the supersymmetry algebra 
\begin{equation} 
\{Q^\pm,Q^\pm\}=2\sqrt2 P^\pm\,,
\qquad \{Q^+,Q^-\}=-4P^\perp\,. 
\end{equation}

In order to express the supercharge in terms of the physical degrees of
freedom, we use constraints which are obtained from the
equations of motion to eliminate the non-dynamical fields
$\chi$ and $A^-$. In light-cone gauge, $A^+=0$, we
reduce the theory dimensionally to two dimensions by setting $\phi=A_2$
and $\partial_2 \rightarrow 0$ for all fields. This yields, from
Eq.~(\ref{supercharges}), 
\begin{equation} 
Q^-=2^{3/4}g\int dx^-
\left(i[\phi,\partial_-\phi]
         +2\psi\psi-\frac{\kappa}{g}
                  \partial_-\phi\right)\frac{1}{\partial_-}\psi\,.
\end{equation}

To discretize the theory we impose periodic boundary conditions on the
boson and fermion fields alike and obtain expansions of the fields 
$\phi_{ij}$ and $\psi_{ij}$ in terms of
discrete momentum modes $A_{ij}(n)$ and $B_{ij}(n)$, respectively.
The positive integers $n$ correspond to discrete longitudinal
momenta $k^+=n\pi/L=nP^+/K$, where $L$ is a longitudinal length
scale, $P^+$ is the total momentum,
and $K$ is a positive integer that determines the resolution.%
\footnote{In DLCQ, $K$ is known as the harmonic resolution~\cite{pb85}.}
The positivity of $n$ guarantees that the number of partons in a 
Fock state is bounded by $K$.  The discrete version of the CS part of the
supercharge is 
\begin{equation} \label{qcs} 
Q^-_{\rm CS}=\left(
\frac{2^{-1/4}\sqrt{L}}{\sqrt{\pi}}\right)
\kappa\sum_{n}\frac{1}{\sqrt{n}}
\left[A^{\dagger}(n)B(n)+B^{\dagger}(n)A(n)\right]\,. 
\end{equation} 
The continuum limit is the limit where $K \rightarrow\infty$. 

Of the two contributions to the supercharge, $Q^-_{\rm SYM}$ and $Q^-_{\rm CS}$,
the former is imaginary and the latter real.
Thus the usual eigenvalue problem 
\begin{eqnarray} \label{EVP}
2P^+P^-|\varphi\rangle&=&\sqrt{2}P^+(Q^-)^2|\varphi\rangle \\
&=& \sqrt{2}P^+(gQ^-_{\rm SYM}+\kappa Q^-_{\rm
CS})^2|\varphi\rangle=M^2_n|\varphi\rangle   \nonumber
\end{eqnarray}
has to be solved by using fully complex methods
to retrieve the mass eigenvalues $M_n$. 

We retain the $Z_2$-symmetry associated with the orientation of the 
large-$N_c$ string of partons in a state~\cite{kutasov93}. 
It gives a sign when the color indices are permuted 
\begin{equation}\label{Z2} 
Z_2 :
A_{ij}(k)\rightarrow -A_{ji}(k)\,, \qquad
      B_{ij}(k)\rightarrow -B_{ji}(k)\,.
\end{equation}
We reduce the numerical effort by using this symmetry to block diagonalize
the Hamiltonian matrix.  Eigenstates will be labeled by the
$Z_2$ sector in which they appear.

It is interesting to note that the pure SYM supercharge is purely 
imaginary.  Consequently, 
the lowest finite-dimensional representation of the superalgebra
is four-dimensional, and there must be an exact four-fold mass degeneracy. 
On the other hand, the full SYM-CS supercharge is complex, and the lowest
complex representation of the superalgebra is two-dimensional. 
Therefore the exact degeneracy only has to be two-fold, 
which is what we find in our numerical results.  


We have converted the mass eigenvalue problem, Eq.~(\ref{EVP}),
to a matrix eigenvalue problem by introducing a discrete basis where the 
longitudinal momentum operator $P^+$ is diagonal.  To obtain the spectrum 
of the SYM-CS theory, we diagonalize the Hamiltonian 
$P^- = (g Q^-_{\rm SYM}+ \kappa Q^-_{\rm CS})^2/\sqrt{2}$.
In the construction we drop the longitudinal zero-momentum modes.  
For some discussion of dynamical and constrained zero modes, see 
the review~\cite{bpp98} and previous work~\cite{alpt98,hpt01}.

The low-energy spectrum can be fit to  $M^2=M^2_{\infty} +  b(1/K)$. 
For a detailed discussion of these fits to the present theory, see 
Ref.~\cite{hpt01}.  The CS term in this theory effectively generates 
a mass proportional to the CS coupling. Therefore, we expect that 
the low-mass states will only have a
few partons, reminiscent of QCD. This is interesting and important because it
stands in stark contrast to ${\cal N}=1$ SYM theory, which is very stringy and
has a large number of low-mass states with a large number of partons. 
\begin{figure}[tbhp]
\begin{center}
\begin{tabular}{c}
\psfig{figure=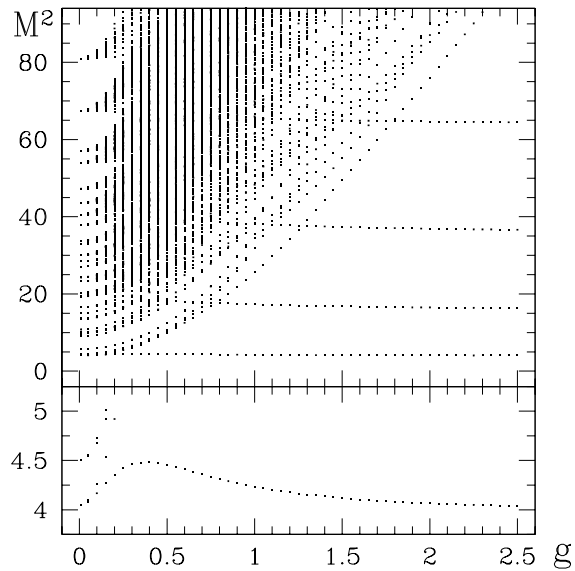,width=7.5cm,angle=0} \\
(a) \\ \vspace{0.1cm} \\
\psfig{figure=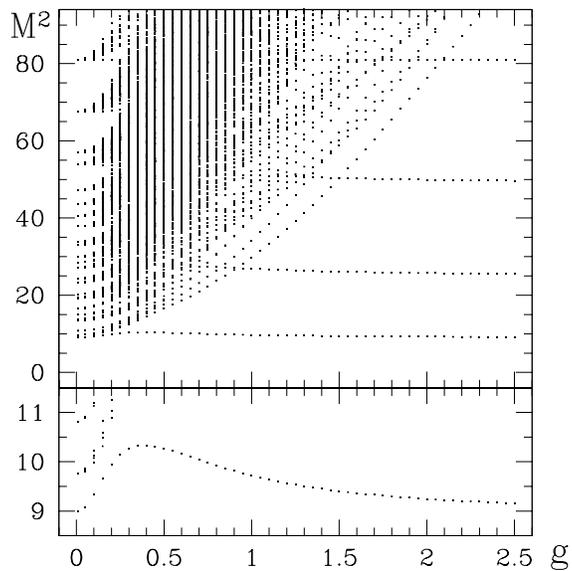,width=7.5cm,angle=0} \\
(b)
\end{tabular}
\end{center}
\caption{Bosonic spectrum at $K=9$ of the two-dimensional theory 
in units of $\kappa^2 N_c$  
as a function of the gauge coupling $g \sqrt{N_c}$ at 
fixed Chern--Simons coupling $\kappa \sqrt{N_c}$ for the
(a) $Z_2$ even sector and (b) $Z_2$ odd sector.
The enlarged plots are of the lowest states in the two
sectors. } 
\label{spec_g0}
\end{figure}

From the structure of the Hamiltonian 
$P^-=(g Q^-_{\rm SYM}+ \kappa Q^-_{\rm CS})^2/\sqrt{2}$
we expect that as a function of $g$ and $\kappa$ the spectrum 
of this theory will grow quadratically in both variables. 
Recall that $Q^-_{\rm SYM}$ is the supercharge of the ${\cal N}=1$ SYM theory 
of adjoint fermions and adjoint bosons, which was studied extensively 
in~\cite{Antonuccio:1998kz}, and $Q^-_{\rm CS}$ is the contribution of the 
CS interaction to the supercharge, given in Eq.~(\ref{qcs}). 
At fixed $g$ as a function of $\kappa$ the spectrum behaves exactly
as expected~\cite{hpt01}. In Fig.~\ref{spec_g0}, we see that at fixed
$\kappa$ the masses of most of the states also grow quadratically in $g$.
There are, however, a number of states 
which behave very differently, 
and it is on these states that we will focus our attention. 

First we need a detailed understanding of the spectrum at $g=0$, because 
we will see that the
theory has a new duality which relates the spectrum at 
$g=0$ to that at $g=\infty$ at fixed $\kappa$. 
The Hamiltonian at $g=0$ is the square of the CS
supercharge, which is simply the Hamiltonian of free fermions and
bosons with mass $\kappa \sqrt{N_c}$. In DLCQ the free $n$-particle 
spectrum at resolution $K$ is given by 
\begin{equation} 
M_n^2(K)=K \kappa^2 N_c \left(\sum_{i=1}^{n-1}\frac{1}{n_i}+
\frac{1}{K-\sum_{i=1}^{n-1}n_i}\right)\,.
\end{equation} 
For example, at $K=3$ the two-parton state has a mass
squared of $M^2=4.5$ in units of $\kappa^2 N_c$, while at 
$K=4$ there are two two-parton
states with eigenvalues  $M^2=4.0$ and  $M^2=5.33$ 
in the same units. These states represent a discrete approximations 
of the two-particle continuum, 
which has its threshold at $M^2=4$ for even 
$K$ and at $M^2=4/(1-1/K^2)$ for K odd.
The threshold for the $n$-particle continuum
is at $M^2=n^2$. It is exactly reproduced at $K=nm$, with $m$ a positive
integer, and approaches this value in the continuum limit otherwise.

We will focus on the states whose energy remains almost constant with
increasing $g\sqrt{N_c}$. These are the states that we classify as
approximate BPS states. They cannot, of course, be true BPS states
because the theory under consideration 
has an ${\cal N}=1$ supersymmetry and can have only
massless BPS states. Rather, these states are a reflection of the massless
BPS states that we found in the pure SYM theory in two 
dimensions~\cite{Antonuccio:1998jg}.%
\footnote{They are also present in the (2+1)-dimensional SYM 
theory~\cite{Haney:2000tk,Antonuccio:1999zu}.} 

In the region where $\kappa/g$ is small, we can understand the connection
of these states to the massless BPS states of pure SYM by doing simple
perturbation theory about the SYM theory.  We re-write the eigenvalue 
equation (\ref{EVP}) by taking out a factor of $g^2$ to obtain
\begin{equation} 
\left(Q^-_{\rm SYM}
+\frac{\kappa}{g}Q^-_{\rm CS}\right)^2|\phi_n\rangle=
{\cal E}_n|\phi_n\rangle\,. 
\end{equation} 
We take $(Q^-_{\rm SYM})^2$ to be the unperturbed Hamiltonian and look for
perturbations of the BPS states $|\psi_n\rangle$ of the pure SYM theory.
Without loss of generality, we assume that they are bosons.
\footnote{There is a set of massless BPS fermions as well.} 
The massless BPS states are approximate
eigenstates of the number operator with ``eigenvalues'' $n=2$ to $K$, and 
we use $n$ to label the $(K-1)$-fold degenerate bound
states.\footnote{This is true to the numerical accuracy of this
calculation, but we have reason to believe that there may be small
additional corrections.}  Of course, the zeroth order energy in the
perturbation expansion vanishes. The important point is that the 
first-order corrections in the degenerate BPS subspace are
determined by the matrix elements 
$\langle\psi_n|\{Q^-_{\rm SYM},Q^-_{\rm CS}\}|\psi_m\rangle$, 
and they vanish as well. From this we see that the
leading order correction to the energy is of order $(\kappa/g)^2$,
and therefore the perturbative expansion for ${\cal E}_n$ can
be written as  
\begin{equation} 
g^2 N_c {\cal E}_n= 
      \kappa^2 N_c {\cal E}_n^{(2)} + 
           {\cal O} \left(\frac{N_c \kappa^3}{g}\right)\,. 
\end{equation} 
Thus the masses of these states are approximately independent of $g$ 
for small $\kappa/g$.  This is exactly what we see in Fig.~\ref{spec_g0}. 
In particular, the detailed plots 
of the lowest states show a $1/g$ convergence of the mass towards the
asymptotic value ${\cal E}_n^{(2)}=n^2$.

We return now to the new duality that these states exhibit. In the 
numerical calculations we find that ${\cal E}_n^{(2)}$ is independent 
of the resolution $K$. Therefore, at large $g^2N_c$ these states are 
exact threshold bound states, again independent of the resolution. 
The unperturbed massless BPS states $|\psi_n\rangle$ 
of the pure SYM theory are, to a good approximation,
diagonal in particle number and have $n$ partons. Therefore, it might not
seem surprising that at $g^2 N_c \rightarrow \infty$ the ``BPS-like" 
states are threshold bound states. They are, however, threshold bound 
states with a special twist. Consider for example the simplest case with 
harmonic resolution $K=3$. The discrete two-particle threshold is at 
$M^2=4.5$ in units of $\kappa^2 N_c$. However, at $g^2 N_c \rightarrow \infty$ 
the ``BPS-like" bound-state mass squared is $M^2=4.0$. This is the true 
threshold, and not the discrete threshold for resolution $K=3$ where the 
calculation was performed. Therefore, at resolution $K=3$ and 
$g^2 N_c \rightarrow \infty$ this bound state is below the discrete 
threshold. From a detailed inspection of the mass  matrix we find that mixing
between the dominant two-particle content and a very small three-particle 
content is essential for
this to occur. The existence of this small three-particle content
is a consequence of a theorem proven in 
Ref.~\cite{Antonuccio:1998kz}, which states
that a pure $n$-parton state cannot exist in two-dimensional 
supersymmetric field theories. 
For even values of the resolution, the discrete threshold
is the correct threshold, and  these ``BPS-like" states are exact
threshold bound states. For odd resolution, the discrete threshold
approaches the correct threshold as we increase the resolution. The
general statement of the new duality is that it relates the masses of these
approximate BPS states at $g^2 N_c= 0$ and at $\infty$ by
\begin{equation}
\lim_{K \rightarrow \infty} {\cal E}_n (g^2 N_c=0) = \lim_{{g \sqrt{N_c}}
\rightarrow \infty} {\cal E}_n (g^2 N_c)\,. 
\end{equation}


To summarize,
we have found that while ${\cal N}=1$ supersymmetric theories cannot have 
true massive BPS states they can have approximate BPS states whose masses
are nearly independent of the Yang--Mills couplings.  The existence of
these states holds out the prospect of allowing one to extrapolate part 
of the massive spectrum of an ${\cal N}=1$ supersymmetric theory into 
the strong-coupling regime in phenomenologically interesting theories.

We found this result in an
analysis of dimensionally reduced SYM-CS theory using the numerical
method of SDLCQ. In SDLCQ we constructed finite
dimensional representations of the superalgebra and therefore a finite
dimensional Hamiltonian which we solved numerically. The SYM-CS theory is
particularly interesting because the CS coupling gives masses to the partons
of this theory without breaking the supersymmetry. This leads to a theory
with QCD-like properties, which is discussed in detail in
Ref.~\cite{hpt01}. 

These approximate BPS states occur because ${\cal N}=1$ SYM theory 
has a set of massless BPS states. Using  ordinary perturbation theory, 
we showed that at small $\kappa/g$ these approximate BPS bound
states are a reflection of the massless BPS bound states of the SYM
theory. Finally, we found that in the limit of infinite coupling $g$
these states are threshold bound states which in turn leads to a duality
relation of the spectrum at $g^2 N_c= 0$ and $\infty$. 

Since these BPS states are also present in the (2+1)-dimensional SYM
theory, we expect to see their reflection  in the (2+1)-dimensional CS
theory.  A study of this theory is underway. 

This work was supported in part by the U.S. Department of Energy. The
authors would like to thank Oleg Lunin for helpful conversations. S.S.P.
would like to acknowledge the Aspen Center for Physics where some of this
work was done.


\end{document}